 \documentstyle[12pt,aaspp]{article}
\begin{document}
\oddsidemargin 16pt
\evensidemargin 16pt
\textwidth 460pt
\topmargin -50pt
\textheight 650pt
\tolerance 1000
\hbadness=10000
\newcommand{\etal}{et al.\ }
\newcommand{\kms}{km~s$^{-1}\ $}
\newcommand{\Lya}{Ly$\alpha\ $}

\title{\Large\bf INTERGALACTIC HYDROGEN CLOUDS AT LOW-REDSHIFT:
    CONNECTIONS TO VOIDS AND DWARF GALAXIES$^{1}$ }
\author{\large J.\ Michael Shull\altaffilmark{1}, John T. Stocke, and
Steve Penton }
\affil{\em Center for Astrophysics and Space Astronomy, \\
      Department of Astrophysical, Planetary, and Atmospheric Sciences, \\
      University of Colorado, CB-389, Boulder CO 80309}
\altaffiltext{2}{also at JILA, University of Colorado and National
Institute of Standards and Technology }
%  \vspace*{0.2cm}
\centerline{To appear in Jan. 1996 {\em Astronomical Journal} }

% \centerline{\today}
% \vspace*{1.5cm}
%  $^{1}$ Based on observations with the NASA/ESA {\it Hubble Space
%  Telescope}, obtained at the Space Telescope Science Institute, which
%  is operated by AURA, Inc., under NASA contract NAS5-26555.
%  \vspace*{0.5cm}

\begin{abstract}
We provide new post-COSTAR data on one sightline (Mrk 421) and updated data
from another (I~Zw~1) from our {\it Hubble Space Telescope} (HST) survey of
intergalactic \Lya clouds located along sightlines to four bright quasars
passing through well-mapped galaxy voids (16,000 \kms pathlength) and
superclusters (18,000 \kms). We report two more definite detections of
low-redshift \Lya clouds in voids: one at 3047 \kms (heliocentric) toward
Mrk~421 and a second just beyond the Local Supercluster at 2861 \kms toward
I~Zw~1, confirming our earlier discovery of \Lya absorption clouds in voids
(Stocke et al. 1995).  We have now identified 10 definite and 1 probable
low-redshift neutral hydrogen absorption clouds toward four targets, a
frequency of approximately one absorber every 3400 \kms above $10^{12.7}$
cm$^{-2}$ column density.  Of these 10 absorption systems, 3 lie within voids;
the probable absorber also lies in a void. Thus, the tendency of \Lya absorbers
to ``avoid the voids'' is not as clear as we found previously.
If the \Lya clouds are approximated as homogeneous spheres of 100 kpc radius,
their masses are $\sim10^9~M_{\odot}$ (about 0.01 times that of bright $L^*$
galaxies) and they are 40 times more numerous, comparable to the density of
dwarf galaxies and of low-mass halos in numerical CDM simulations.  The \Lya
clouds contribute a fraction $\Omega_{cl} \approx 0.003 h_{75}^{-1}$ to the
closure density of the universe, comparable to that of luminous matter. These
clouds probably require a substantial amount of non-baryonic dark matter for
gravitational binding.  They may represent extended haloes of low-mass
protogalaxies which have not experienced significant star formation or low-mass
dwarf galaxies whose star formation ceased long ago, but blew out significant
gaseous material.
\end{abstract}

\newpage

\section{INTRODUCTION}

Over the past two decades, redshift surveys of the nearby universe (Geller \&
Huchra 1989;  Fairall et al. 1990)
have mapped out a highly inhomogeneous galaxy distribution, with
large ``voids'' bounded by sheetlike superclusters.  A goal of studies of the
content of these voids (Sanduleak \& Pesch 1987; Szomuru et al. 1994; Weinberg
et al. 1991; Strauss \& Huchra 1988) is to understand how galaxy formation and
subsequent interactions fit into a cosmological framework. With the {\it Hubble
Space Telescope} (HST), we have undertaken a spectral search for low-redshift
neutral hydrogen (\Lya) absorption clouds along sightlines to bright quasars
behind well-mapped voids and superclusters. In our earlier work from Cycle 2
(Stocke et al. 1995, hereafter Paper I), on the first three sightlines, we
detected 8 definite \Lya absorption lines, ranging in equivalent width from 26
to 240 m\AA.  Seven of these absorbers were located in supercluster galaxy
structures. One absorber, in the sightline toward Mrk~501, was located in a
void, more than 5.9 Mpc from the nearest bright galaxy.

This discovery demonstrated that the voids are not entirely empty of gaseous
matter, but the statistics were low.  In fact, a number of H~I galaxies, IRAS
galaxies, and emission-line galaxies have been found within the boundaries of
the Bootes void (Tifft et al. 1986; Weistrop \& Downes 1988; Dey, Strauss, \&
Huchra 1990; Szomoru et al. 1993), the Pegasus void (Fairall et al. 1990), and
the Pisces-Perseus void (Henning 1992). However, the filamentary structure of
galaxies inside the Bootes and Pegasus voids is similar to that found in the
voids between the Local Supercluster and the ``Great Wall'' in the CfA survey
(Huchra et al. 1993; see also our Figure 1).  Because the total extent of the
voids is somewhat uncertain at the distance of Bootes, we chose our targets
toward more local voids, with better-mapped galaxy distributions. For example,
one \Lya cloud in the direction of Mrk~501 reported in Paper I is found within
a filamentary structure of galaxies separating two large voids whose total size
is comparable to the Bootes void.

In this paper, we report the results of additional HST Cycle 4 observations of
I~Zw~1 and Mrk~421. Toward I~Zw~1, we are now able to upgrade a ``probable''
void absorber to ``definite'' status, and we have detected another \Lya
absorber in a void toward the fourth target, Mrk 421. We update the
distribution of \Lya clouds toward all four sightlines, analyze the physical
parameters of these \Lya clouds, and make an estimate of their contribution to
the baryonic mass density of the Universe. We conclude with a discussion of a
possible connection of \Lya clouds with dwarf galaxies and the implications of
our detections for cosmology and galaxy formation.

\section{OBSERVATIONS}

The spectra of our first three targets (Mrk~501, Mrk~335, I~Zw~1) were taken
with the Goddard High Resolution Spectrograph (GHRS) aboard the HST, using the
G160M grating with pre-COSTAR optics.  These data and their interpretation are
discussed in detail by Stocke et al. (1995, hereafter Paper I). Since that
paper, we have re-observed I~Zw~1 (6.53 hr on 6 February 1995) and obtained a
new spectrum (4.35 hr on 1 February 1995) on our fourth target, Mrk 421. I~Zw~1
(0050+1225) is a bright quasar ($V = 14.0$, $cz = 18,300$ \kms) located at
$\ell = 123.75^{\rm o}$ and $b = -50.2^{\rm o}$, while Mrk 421 is a bright BL
Lac object ($V = 12.9$, $cz = 8880$ \kms) located at $\ell = 179.8^{\rm o}$ and
$b = +65.0^{\rm o}$.  Both observations used the G160M grating and the COSTAR
enhancement of the GHRS optics to obtain 40 \kms resolution in the interval
1222 -- 1259 \AA. These wavelengths correspond to redshifted velocities $cz
\approx 1500 - 10,500$ \kms extending roughly from the local supercluster to
the ``Great Wall'' (Geller \& Huchra 1989). In this paper we assume a Hubble
constant $H_0 = (75~{\rm km~s}^{-1}~{\rm Mpc}^{-1}) h_{75}$. The four QSO
targets (Mrk~501, Mrk~335, I~Zw~1, and Mrk~421) were chosen because of their
brightness ($V \leq 14$) and their location behind well-studied galaxy
distributions with foreground voids considerably emptier than the Bootes Void
(Sanduleak \& Pesch 1987; Szomoru et al. 1994; Weinberg et al. 1991; Strauss \&
Huchra 1988).

Our data reduction was identical to that in Paper I. All spectra were taken
through the $2''$ large science aperture using the standard quarter-diode
sub-stepping pattern to yield pixels of 0.018 \AA\ in the FP-split mode. We
reduced our spectra with the STSDAS spectral reduction package within IRAF, and
the wavelength scale was determined by assuming that the Galactic S~II
absorption features at 1250.584 \AA\ and 1253.811 \AA\ lie at zero velocity in
the local standard of rest (LSR). The statistical significance of the
absorption features was determined from measured noise in the continuum and
from uncertainties in continuum placement.  The status of ``definite''
absorber was given to features greater than $4 \sigma$ significance;
``probable'' detections were those of 3--4 $\sigma$ significance.

In Figure 1 we show the co-added spectrum of I~Zw~1 (6.53 hrs of new data added
to 7.21 hrs of previous data taken on 8 September 1993). We have definite
detections of five spectral features:  three \Lya lines and two Galactic S~II
absorbers.  By combining the two spectra, we confirm our previous definite
detections of \Lya absorbers at heliocentric velocities $1617 \pm 5$ \kms and
$5130 \pm 12$ \kms, and we upgrade the ``probable'' detection of the $2861 \pm
9$ \kms absorber to definite status. The new spectrum of Mrk~421 is shown in
Figure 2.  We detect a single \Lya absorber at high significance at
velocity $ 3046 \pm 2$ \kms and two Galactic S~II lines. Table 1
lists the wavelengths, heliocentric velocities, equivalent widths, and
significances of the features.

The two new absorbers toward I~Zw~1 and Mrk~421 both lie within voids,
confirming our previous discovery of \Lya clouds in voids. The galaxy
distributions are shown in Figure 3, based on the CfA merged galaxy redshift
catalog (Huchra et al. 1993).  Because the Mrk~501 sightline lies near the
eastern edge of the CfA survey, we obtained 20 additional redshifts at RA
greater than 17 hr from the ongoing survey to verify that the voids in front of
Mrk~501 extend well beyond 17 hr (Huchra, private communication).  The galaxy
distributions toward Mrk~501 and Mrk~421 are shown at somewhat higher
declinations ($\delta = 39^{\rm o} \pm 4^{\rm o}$) than those usually displayed
by the CfA group (deLapparent, Geller, \& Huchra 1986; Peebles 1993). These
sightlines traverse comparable portions of ``supercluster'' and ``void'',
determined according to the ``wavelet analysis'' method (Slezak, de Lapparent,
\& Bijaoui 1993).  We believe that our identifications of void and supercluster
regions are accurate to $\pm500$ \kms.

We now compare our new results and those of Paper I and update the statistics
based on the new data.  Over the four sightlines, the pathlength ($c \Delta z$)
through voids totals 16,000 \kms, while that through superclusters totals
18,000 \kms. We have now detected a total of 10 definite \Lya absorption lines
at a significance above $4 \sigma$ (Table 2). The \Lya equivalent widths
detected with the GHRS range from 26 to 240 m\AA, corresponding to column
densities N(H~I) between $10^{12.7}$ and $10^{14}$ cm$^{-2}$ for Doppler
parameter $b \approx 30$ km~s$^{-1}$.  The frequency of absorbers is
approximately one every 3400 \kms of pathlength, above $10^{12.7}$ cm$^{-2}$
column density, comparable to that estimated from GHRS observations of other
AGN (Weymann 1993; Savage, Sembach, \& Lu 1995).  These numbers are
approximate, however, since proper treatments of the GHRS sensitivity function
and the H~I column density distribution have not yet been performed.

Three clouds lie within voids (one each toward Mrk~501, I~Zw~1, and Mrk~421)
and seven are associated with supercluster walls, including three in the Local
Supercluster. Expressed as heliocentric velocities, the three definite void
absorbers lie at 3046 \kms toward Mrk~421, at 7740 \kms toward Mrk~501, and at
2861 \kms toward I~Zw~1. The nearest known galaxies lie 4.3, 5.9, and 1.7
$h_{75}^{-1}$ Mpc away from these clouds, respectively (Table 2). In the case
of absorber B toward I~Zw~1, the nearest galaxy (A0054+1005 at $cz = 2890$ \kms
and $1.7 h_{75}^{-1}$ Mpc distance) is likely to be part of the Local
Supercluster.  We had classified the region surrounding the 2861 \kms absorber
as a void prior to the HST/GHRS observations, and we believe that this absorber
lies just beyond the Local Supercluster in that direction. The distances to
other bright galaxy neighbors are considerably larger. At somewhat lower
significance (3--4 $\sigma$) we find one more probable \Lya absorber at 6000
\kms which also lies in a void toward Mrk~501.

Therefore, with combined pathlength 47\% through voids and 53\% through
superclusters, we find that 3 of 10 definite \Lya absorbers lie within voids.
These statistics are small, and the addition of the fourth QSO sightline has
shifted the fraction of \Lya absorbers within voids slightly higher than that
(1 out of 8) quoted in Paper I. According to binomial statistics for a uniform
distribution of absorbers, the observed (3 out of 10) situation could occur
with 15\% chance probability (19\% when one includes the 1 additional
``possible'' absorber which also lies in a void). Therefore, the tendency of
\Lya clouds to lie in superclusters cannot be stated as definitively as in
Paper I. The range of distances to nearest-neighbor bright galaxies for the
three void absorbers (1.7 -- 5.9$h_{75}^{-1}$ Mpc) is comparable to that for
the seven non-void absorbers (0.45 -- 5.9$h_{75}^{-1}$ Mpc).  However, the
nearest neighbor to the possible void absorber toward Mrk 501 lies $10.5
h_{75}^{-1}$ Mpc away, and the distribution of distances to the nearest
galaxies to the void and non-void absorbers shows some differences inside and
outside 1 Mpc (Paper I). For example, four of the seven non-void absorbers have
neighbor galaxies within 1 Mpc. Evidently the discovery of more absorbers
toward AGN behind well-mapped galaxy distributions is needed to settle the
issue of how well \Lya absorbers trace the large-scale structure of galaxies.

These nearest bright galaxies are too far from the \Lya clouds to be physically
associated in most models. Pencil-beam optical and 21-cm surveys of the area of
sky surrounding Mrk~501 find no galaxy at comparable redshift to the 7530 \kms
void absorption system within $100 h_{75}^{-1}$ kpc with absolute magnitude $B
\leq -16$ and no object with H~I mass $\geq (7 \times
10^8~M_{\odot})~h_{75}^{-2}$ within 500 $h_{75}^{-1}$ kpc (Paper I). Thus,
neither a faint optical galaxy nor a gas-rich galaxy is present close to this
\Lya cloud. However, as we shall suggest in the next section, it is probable
that the low-redshift \Lya absorbers may be associated with dwarf galaxies.

\section{CHARACTERISTICS OF THE CLOUDS }

We now derive the physical characteristics of the H~I clouds. If the clouds are
exposed to metagalactic photoionizing radiation of specific intensity $J_{912}
= (10^{-23}$ ergs~cm$^{-2}$ s$^{-1}$ sr$^{-1}$ Hz$^{-1})J_{-23}$ at 912 \AA\
and (energy) spectral index $\alpha_s \approx 1.5$, the hydrogen ionization
correction can be large, but derivable from the assumptions of photoionization
equilibrium and cloud geometry. The void \Lya lines toward Mrk~501, I~Zw~1, and
Mrk~421 have equivalent widths of 48 m\AA, 65 m\AA, and 92 m\AA\ respectively,
corresponding to column densities N(H~I) $\approx 1-3 \times 10^{13}$ cm$^{-2}$
for Doppler-broadening parameter $b \approx 30$ \kms. Other low-redshift \Lya
clouds have H~I columns up to $10^{14}$ cm$^{-2}$.  Since the total mass of
photoionized homogeneous \Lya clouds is probably dominated by the upper end of
the distribution, we will hereafter scale to a characteristic column density
N(H~I) = $(10^{14}$~cm$^{-2})N_{14}$.

To make further progress, we need an estimate for the cloud sizes. However,
unlike the situation with high-redshift \Lya clouds, we lack definitive
information on the size of the low-$z$ absorbers.  From the absorber
coincidences and anti-coincidences toward the double-quasar Q1343+266A,B at $z
= 2.03$ (Bechtold et al. 1994; Dinshaw et al. 1994) the radii of \Lya absorbers
at $z = 1.79- 2.03$ have been estimated to lie in the range $50 \leq R \leq
370~h_{75}^{-1}$ kpc, with a median value of $120~h_{75}^{-1}$ kpc. At low
redshift, no such measurements have been performed. Sizes much smaller than 100
kpc would require unacceptably large space densities (Weymann 1993; Maloney
1993). The only physical information on low-$z$ cloud size comes from studies
(Lanzetta et al. 1995) of 46 galaxies in the fields of HST/FOS \Lya absorbers.
Each of five galaxies within $90~h_{75}^{-1}$ kpc corresponds to a \Lya
absorber, whereas the frequency is only five of ten galaxies between
$90~h_{75}^{-1}$ and $200~h_{75}^{-1}$ kpc and one of nine galaxies beyond
$200~h_{75}^{-1}$ kpc. Lanzetta et al (1995) also found an anti-correlation
between \Lya equivalent width and (\Lya - galaxy) impact parameter.  However,
as we showed in Paper I, the trend for lower-column absorbers to have larger
impact parameters does not appear to extend beyond 100--200 kpc for the
lower-column H~I clouds detected by HST/GHRS.  Taking all this evidence into
account, we adopt a characteristic radius $R = (100~{\rm kpc}) R_{100}$ and
carry along the scaling parameter $R_{100}$ to demonstrate the sensitivity of
some quantities to this uncertain size.

The precise distribution of density in these \Lya clouds is uncertain, and mass
estimates based on photoionization corrections are imprecise. To obtain a
first-order estimate, we assume that the clouds are homogeneous spheres of
radius $R = (100~{\rm kpc})R_{100}$ and temperature $T = (10^{4.3}~{\rm
K})T_{4.3}$. In photoionization equilibrium, the neutral hydrogen density,
$n_{HI}$, is
\begin{equation}
    n_{HI} = \frac {n_e~n_{HII}~\alpha_{H}^{(1)} } { \Gamma_H }
       = (10.8) n_H^2~T_{4.3}^{-0.726}~J_{-23}^{-1}
          \left[ \frac {\alpha_s+3}{4.5} \right] \; ,
\end{equation}
where we adopt an electron density $n_e = 1.157 n_H$ (for helium abundance $Y =
0.239$) and assume a radiative recombination rate coefficient $\alpha_H^{(1)} =
(2.48 \times 10^{-13}~{\rm cm}^3~{\rm s}^{-1}) T_{4.3}^{-0.726}$ and hydrogen
photoionization rate $\Gamma_H = (2.66 \times 10^{-14}~{\rm s}^{-1})
[4.5/(\alpha_s+3)] J_{-23}$.  If the mean column density is taken to be
N(H~I) $\approx n_{HI} R$, then the total hydrogen density $n_H$, neutral
fraction $x_{HI} = n_{HI}/n_H$, and cloud mass $M_{cl}$ are,
\begin{eqnarray}
n_H    &=& (5.5 \times 10^{-6}~{\rm cm}^{-3}) ~J_{-23}^{1/2} ~T_{4.3}^{0.363}
          ~ N_{14}^{1/2} ~R_{100}^{-1/2} ~\left[ \frac{\alpha_s+3}{4.5}
            \right]^{-1/2} \; , \\
x_{HI} &=& (5.9 \times 10^{-5})~ J_{-23}^{-1/2}~ T_{4.3}^{-0.363}~
              N_{14}^{1/2} ~R_{100}^{-1/2} ~\left[ \frac{\alpha_s+3}{4.5}
              \right]^{1/2} \; , \\
M_{cl} &=& (7.4 \times 10^8~M_{\odot}) ~J_{-23}^{1/2} ~T_{4.3}^{0.363}~
        N_{14}^{1/2} ~R_{100}^{5/2} ~\left[ \frac{\alpha_s+3}{4.5}
         \right]^{-1/2} \; .
\end{eqnarray}
{}From the observed absorber frequency of one cloud with N(H~I) $\geq 10^{13}$
cm$^{-2}$ every 3500 \kms (Weymann 1993),
one infers a redshift frequency $dN/dz = \phi_0 (\pi
R^2)(c/H_0) \approx 86$.  The local space density of the low-redshift \Lya
systems is therefore
\begin{equation}
   \phi_0 = (0.68~{\rm Mpc}^{-3}) R_{100}^{-2} h_{75} \; ,
\end{equation}
approximately 40 times that of bright galaxies, $L^* = (9.1 \times
10^{9}~L_{\odot}) h_{75}^{-2}$, with luminosity function $\phi(L)~dL \equiv
\phi^* (L/L^*)^{\alpha} \exp(-L/L^*)(dL/L^*)$. The recent CfA survey (Marzke,
Huchra, \& Geller 1994) gave values $\phi^* = (0.017 \pm 0.004~{\rm Mpc}^{-3})
h_{75}^{3}$ and $\alpha = -1.0 \pm 0.2$, while other determinations (Sandage,
Binggeli, \& Tammann 1985; Impey, Bothun, \& Malin 1988) of $\phi(L)$ find a
faint-end slope $\alpha \approx -1.3$ or even steeper in the Virgo Cluster.
However, statistics on the faint-end luminosity function are poor, and values
of the slope ($\alpha$) and normalization ($\phi^*$) trade off against one
another.  For example, Marzke et al. (1994) noted a factor-of-three excess at
the faint end, relative to their ($\alpha = -1.0$) fit, and claimed that this
may be due to Sm-Im dwarf galaxies with $\alpha = -1.87$.

We are now in a position to compare the space density of the \Lya clouds to
that of galactic consituents of the Universe.  The volume filling factor of the
\Lya clouds is
\begin{equation}
   f = \left( \frac {4RH_0}{3c} \right) \left( \frac {dN}{dz} \right)
        = (0.0029) R_{100} h_{75}  \; .
\end{equation}
If $L^*$ galaxies have a mass-to-light ratio (Persic \& Salucci 1992) of
$9h_{75}~M_{\odot}/L_{\odot}$ within the Holmberg radius, the \Lya cloud masses
correspond to $0.01 m^*$, where $m^* \approx (8.2 \times 10^{10}~M_{\odot})
h_{75}^{-1}$. For a faint-end slope $\alpha = -1.3$, we estimate that galaxies
with $0.003 L^* \leq L \leq 0.03 L^*$ have a space density $0.16~{\rm Mpc}^{-3}
h_{75}^3$, four times less than that estimated for ($R = 100$ kpc) \Lya
absorbers. Given the uncertainties in the luminosity function and \Lya cloud
size, these space densities are in fair agreement.  They are also comparable to
those of low-mass halos found in estimates using the Press-Schechter formalism
and in recent numerical CDM simulations (Efstathiou 1995), which predict a
comoving space density $N(>M_h) = 1.7 h_{75}^3$ Mpc$^{-3}$ of halos with mass
greater than $M_h \approx 5 \times 10^{9}~h_{75}^{-1}~M_{\odot}$. That
particular simulation found $N(>M_h) \propto M_h^{-1}$ at low masses,
appropriate for hierarchical clustering models in which the power spectrum of
mass fluctuations $P(k) \propto k^{-3}$. An unanswered question from these
simulations is the fraction of low-mass halos that undergo star formation at
some epoch.

Many surveys of the faint end of the luminosity function (Sandage et al. 1985;
Impey et al. 1998; Marzke et al. 1994) suggest a large space density of dwarf
galaxies, some of which are difficult to detect owing to their low surface
brightness.  The \Lya absorbers could be the halos of the faint blue dwarf
galaxies (Tyson 1988;  Cowie et al. 1988) that have faded by the present epoch
(Babul \& Rees 1992; Salpeter 1993) or low-mass gas clouds whose cores have not
yet undergone significant star formation. Our previous failure to detect any
dwarf galaxies down to $M_B = -16$ at the position of the void absorber toward
Mrk~501 (Stocke et al. 1995) accentuates the need for deeper imaging of these
fields in search of dwarf galaxies down to $M_B \approx -13$.   In the only
such search conducted thus far, Rauch, Weymann, \& Morris (1995) failed to
find any galaxied with $M_B \leq -13$ near the 3C~273 sight line, including low
surface brightness objects with $\Sigma_B \leq 26.5$ mag arcsec$^{-2}$. Optical
and 21 cm searches for such dwarfs are possible only for very local \Lya
clouds, since many of the \Lya clouds (Bahcall et al. 1991; Morris et al. 1991)
disovered by HST are too distant to set sensitive limits on the presence of
dwarf galaxies. H~I (21-cm) searches at $cz \leq 10,000$ \kms are
straightforward, but studies of more distant gas require increasingly large
amounts of observing time.

Similarly,
spectra of the low-$z$  \Lya clouds in the ultraviolet resonance lines of C~IV,
C~III, and C~II could set useful limits on possible star formation.  From
equation (2), the column density of carbon in all ion stages should be N(C)
$\approx (8 \times 10^{12}$~cm$^{-2})R_{100}^{1/2} N_{14}^{1/2} J_{-23}^{1/2}$
at 0.01 solar metallicity. If 10\% of the carbon is in C~IV, absorption lines
at 1549 \AA\ should be present at a very weak level (1--10 m\AA\ equivalent
width). Lines from C~II (1335 \AA) or C~III (977 \AA) might set even better
metallicity limits if carbon exists in lower ionization states owing to a less
intense or softer radiation field.

We can obtain an estimate of the fractional contribution, $\Omega_{cl} = \phi_0
M_{cl}/ \rho_{cr}$, of low-redshift \Lya clouds to the critical (closure)
density, $\rho_{cr} = (1.06 \times 10^{-29}~{\rm g~cm}^{-3}) h_{75}^2$, of the
universe from equations (4) and (5).  The value,
\begin{equation}
   \Omega_{cl} = (0.0033)~J_{-23}^{1/2}~ T_{4.3}^{0.363} ~N_{14}^{1/2}
  ~ R_{100}^{1/2} ~h_{75}^{-1} ~\left[ \frac{\alpha_s+3}{4.5}
       \right]^{-1/2} \; ,
\end{equation}
is comparable to that ($\Omega_* \approx 0.004$) of optically-luminous matter
(Peebles 1993) and is 15\% of the total baryonic limit, $\Omega_b = (0.022 \pm
0.004) h_{75}^{-2}$, inferred (Walker et al. 1991) from the constraints of Big
Bang nucleosynthesis. Thus, these clouds have overdensities $\sim50
J_{-23}^{1/2} N_{14}^{1/2} R_{100}^{-1/2} T_{4.3}^{0.363}$ with respect to the
smooth baryonic background density, $\Omega_b \rho_{cr}$.

We note that our estimates for the low-redshift absorbers are uncertain, since
they depend on poorly known geometric parameters of the clouds.  For example,
we do not know the distributions of cloud size and N(H~I), assumed here to be
constant at 100 kpc and $10^{14}$ cm$^{-2}$ respectively.  We note that, while
cloud mass scales as $R_s^{5/2}$, the total mass density and value of
$\Omega_{cl}$ are less sensitive ($R_s^{1/2}$) to assumptions of cloud size. In
addition, the distribution of matter within the low-redshift absorbers might
also be inhomogeneous, either with a radial distribution of density or clumps.
Each of these distributions would accentuate, through recombinations, the
regions with highest density.  One could also relax the assumption of spherical
clouds.  By generalizing the spherical assumption to disk geometries, Madau \&
Shull (1996) showed that the value of $\Omega_{cl}$ is reduced by a factor
$\langle a~\cos \theta \rangle^{-1/2} \approx (2/a)^{1/2}$, where $a > 1$ is
the disk aspect ratio and $\theta$ is their viewing angle. So, for example, a
ratio $a = 10$ would reduce $\Omega_{cl}$ by a factor 2.2.

\section{DISCUSSION}

In the proto-galaxy scenario, the low-$z$ \Lya clouds represent parcels of gas
and dark matter that have recently ($z < 0.5$) turned around from the general
Hubble expansion and now have mean overdensities $\sim 50-200$ times that of
the background IGM. To be consistent with the frequency of absorbers, these
parcels must have a space density comparable to that of dwarf galaxies.  The
hydrogen clouds around these galaxies may only recently have become detectable
in \Lya absorption because of the rapid drop in the ionizing radiation field at
low redshift. Whether these gaseous envelopes could retain their integrity
following encounters with passing massive galaxies would depend on the
distances of these galaxies and on the gas dynamical outcome of the encounter.
No encounters may have occurred with the void absorbers toward Mrk 501 and Mrk
421, since neighboring galaxies from the supercluster walls moving over 10 Gyr
with $\leq500$ \kms peculiar velocity would not be able to traverse the void.

In the faded dwarf galaxy scenario (Babul \& Rees 1992; Efstathiou 1994), the
low-redshift clouds are the extended gaseous halos of the faint blue dwarf
galaxies detected in deep imaging surveys. Although we have modeled these
objects as homogeneous spheres, it is likely that the cloud cores have
collapsed and formed stars, whose winds and supernovae blow matter out into the
extended (100 kpc) envelopes required by the frequency, $dN/dz$, of \Lya
absorbers. The fact that some \Lya clouds reside in regions devoid of bright
galaxies is consistent with the observed low amplitude of angular correlation
found (Efstathiou et al. 1991) for the faint blue galaxies. The scenario
predicts that very low luminosity galaxies, similar to the Local Group dwarfs,
should be found in deep (B$_{\rm lim} \approx 23$) ground-based optical
images near the low-redshift \Lya clouds.  In a deep imaging search for faint
galaxies toward 3C~273 near the local \Lya clouds in the Virgo Cluster, Morris
et al. (1993) and Rauch, Weymann, \& Morris (1995) found no dwarf galaxies to
impressively faint limits ($M_B \leq -13$ and limiting surface brightness
levels of $\sim 26.5$ mag~arcsec$^{-2}$).  Similarly sensitive searches are
required for cloud sample.  Thus, H~I (21 cm) observations and deep optical
imaging are in progress (Carilli et al. 1995) to search for accompanying faint
galaxies.

Finally, it is worth speculating on possible dark-matter content.
Since the Jeans mass, $M_J = (2 \times 10^{11}~M_{\odot})R_{100}^{1/4}
J_{-23}^{-1/4} N_{14}^{-1/4} T_{4.3}^{1.32}$, safely exceeds the baryonic mass,
the clouds are not gravitationally unstable. However, in order to bind them
gravitationally, we require that $(2kT/m_H) \approx (G\beta M_{cl}/R)$, where
the ratio of total matter to baryonic matter is
\begin{equation}
     \beta \equiv M_{tot}/M_{cl} = (10)~ R_{100}^{-3/2}~ N_{14}^{-1/2}~
        J_{-23}^{-1/2}~T_{4.3}^{0.636}   \; .
\end{equation}
Thus, as in other large self-gravitating systems, the low-redshift \Lya clouds
could contain substantial amounts of dark matter, unless the clouds are quite
large ($R \approx 400$ kpc) as suggested for some clouds at higher redshift
(Dinshaw et al. 1995). Since all baryonic matter has been included in the
ionization correction (equation [1]) this dark matter would need to be
non-baryonic.

\vspace{1cm}

Support for this work was provided by HST Guest Observer grant
(GO-3584.01-91A) and the NASA Astrophysical Theory program (NAGW-766) at
the University of Colorado.  We thank Mark Giroux and Phil Maloney
for helpful discussions.

\newpage

\begin{table}[h]
  \begin{center}
  \caption{~~~DETECTED ABSORPTION LINES$^a$ TOWARD I~Zw~1 AND Mrk~421 }
  \vskip 10pt
  \begin{tabular}{lcccccl}
  \hline \hline \\
Target & System & Wavelength & Velocity & $W_{\lambda}$ &
         Significance  & Identification \\
       & Designation & (\AA) & (\kms)   & (m\AA)  &  ($\sigma$) &   \\
 \hline
I~Zw~1  & A     & 1222.23 & $1617 \pm 5$  & $120 \pm 37$ & 7.1 & Ly$\alpha$  \\
I~Zw~1  & B     & 1227.27 & $2861 \pm 9$  & $84 \pm 40$  & 4.7 & Ly$\alpha$  \\
I~Zw~1  & C     & 1236.47 & $5130 \pm 12$ & $84 \pm 47$  & 4.9 & Ly$\alpha$  \\
I~Zw~1  &       & 1250.56 &  .....        & $109 \pm 44$ & 8.9 & S~II Galactic
\\
I~Zw~1  &       & 1253.84 &  .....        & $132 \pm 37$ & 10.5& S~II Galactic
\\
        &       &         &               &              &     &
\\
Mrk~421 & A     & 1228.02 & $3046 \pm 2$  & $92 \pm 10$  & 24  & Ly$\alpha$
\\
Mrk~421 &       & 1250.55 &  .....        & $122 \pm 11$ & 35  & S~II Galactic
\\
Mrk~421 &       & 1253.79 &  .....        & $195 \pm 10$ & 61  & S~II Galactic
\\
  \hline
  \end{tabular}
  \end{center}

 $^a$ Features detected above $4 \sigma$ significance.  Velocities are
heliocentric, assuming that the S~II lines are at the Galactic LSR.
\end{table}

%  \newpage

\begin{table}[h]
  \begin{center}
  \caption{~~~NEAREST GALAXY NEIGHBORS TO \Lya CLOUDS }
  \vskip 10pt
  \begin{tabular}{lcclccc}
  \hline \hline \\
Target  & System      & Absorber     & Nearest & Galaxy  & $\delta \Theta$ &
                                                         Distance$^b$ \\
        & Designation & Velocity$^a$ & Galaxy  & Velocity$^a$ & (degrees) &
                                                        ($h_{75}^{-1}$ Mpc) \\
    &   &  (km~s$^{-1}$) &  & (km~s$^{-1}$)&       &           \\
  \hline
Mrk~335 &  A & 1970  &  NGC~7817    &  2520 & 0.80 & 0.67  \\
Mrk~335 &  B & 2290  &  NGC~7817    &  2520 & 0.80 & 0.45  \\
Mrk~335 &  C & 4270  &  00008+2150  &  4670 & 2.00 & 2.2    \\
Mrk~335 &  D & 6280  &  00036+1928  &  6150 & 0.46 & 0.85   \\
Mrk~501 &  A & 4660  &  16510+3927  &  4830 & 0.45 & 0.51  \\
Mrk~501 &  B:& 6000  &  IC~1221     &  5690 & 7.6  & 10.5  \\
Mrk~501 &  C & 7530  &  1709+3941   &  7810 & 3.3  &  5.9   \\
I~Zw~1  &  A & 1617  &  NGC~63      &  1300 & 8.8  &  3.3 \\
I~Zw~1  &  B & 2861  &  A0054+1005  &  2890 & 2.5  &  1.7  \\
I~Zw~1  &  C & 5130  &  00522+1325  &  5560 & 1.0  &  1.3  \\
Mrk~421 &  A & 3046  &  11267+3508  &  2540 & 5.6  &  4.3  \\
 \hline
 \end{tabular}
 \end{center}

 $^a$ Heliocentric velocities

 $^b$ Three-dimensional distance from \Lya absorber to galaxy,
   using a retarded Hubble flow model (Paper I)

\end{table}

\newpage

\section*{Figure Captions}
\begin{itemize}

\item[{\em Figure 1}:]  HST/GHRS spectrum of I~Zw~1 taken with G160M grating;
figure shows combined data from pre-COSTAR (7.21 hr) and post-COSTAR (6.53 hr)
exposures. Three definite \Lya absorbers are detected at wavelengths 1222.2,
1227.3, and 1236.5 \AA.  The second absorber at heliocentric velocity $2861
\pm 9$ \kms and equivalent width $84 \pm 40$ m\AA\ lies within the void of
foreground galaxies. Lines at 1250.59 and 1253.81~\AA\ are Galactic
interstellar S~II, and the broad emission line is C~III
$\lambda1175$ from the QSO.

\item[{\em Figure 2}:]  HST/GHRS spectrum of Mrk~421 taken with G160M grating
and post-COSTAR optics.  A \Lya absorber at $3046 \pm 2$ \kms and equivalent
width $92 \pm 10$ m\AA\ lies within the void of foreground galaxies. Lines at
1250.59 and 1253.81 \AA\ are Galactic interstellar S~II. A second arrow marks
the expected (1252 \AA) position of redshifted Ly$\alpha$ from the BL Lac
object ($z_{\rm em} = 0.0298$).  The line at 1258.5 \AA\ is unidentified; its
wavelength is 1 \AA\ shortward of S~II 1259.52 \AA, but the wavelength scale at
the long end is uncertain.

\item[{\em Figure 3}:]  ``Pie diagram'' distributions in galactocentric
recession velocity and right ascension of the \Lya absorbers toward our four
target AGN.  The AGN are shown as the farthest symbols along the sightlines,
and the \Lya absorbers are shown as circles (the 3 definite and 1 probable
absorber in voids have an inscribed ``X''). (a) 3 absorbers toward Mrk 501 (at
4880, 6220, and 7740 \kms heliocentric velocity) and 1 toward Mrk~421 (at 3046
\kms); (b) 3 absorbers toward I~Zw~1 (at 1617, 2861, and 5130 \kms) and 4
toward Mrk~335 (at 2185, 2520, 4490, and 6490 \kms).

\end{itemize}
\end{document}